**Textured ceramics with controlled grain size and orientation**

Rohit Pratyush Behera[1], Syafiq Bin Senin Muhammad[1], Marcus He Jiaxuan[2], Hortense Le Ferrand*[1,2]

[1]School of Mechanical and Aerospace Engineering, Nanyang Technological University, 50 Nanyang avenue, Singapore 639798

[2]School of Materials Science and Engineering, Nanyang Technological University, 50 Nanyang avenue, Singapore 639798

*Corresponding email: hortense@ntu.edu.sg

*Abstract. Texture in dense or porous ceramics can enhance their functional and structural properties. Current methods for texturation employ anisotropic particles as starting powders and processes that drive their orientation into specific directions. Using ultra-low magnetic fields combined with slip casting, it is possible to purposely orient magnetically responsive particles in any direction. When those particles are suspended with nanoparticles, templated grain growth occurs during sintering to yield a textured ceramic. Yet, the final grains are usually micrometric, leading weak mechanical properties. Here, we explore how to tune the grain orientation size using magnetic slip casting and templated grain growth. Our strategy consists in changing the size of the anisotropic powders and ensuring that magnetic alignment and densification occurs. The obtained ceramics featured a large range of grain anisotropy, with submicrometric thickness and anisotropic properties. Textured ceramics with tunable grains dimensions and porosity are promising for filtering, biomedical or composite applications.*

**1| INTRODUCTION**

Textured ceramics have the potential to enhance adsorption and catalysis efficiency in porous filters [1], diffusion in solid-state batteries [2], cell growth and biointegration in implants [3], to serve as preforms for anisotropic reinforced composites with combined strength and toughness [4], transparency [5], and to create dense ceramics with outstanding mechanical, electromechanical and/or thermal properties [6–10]. Manufacturing methods of bulk textured porous or dense ceramics for those applications employ anisotropic porogens [3], directional freeze-casting [11], tape-casting [12], high magnetic fields [13] or unidirectional pressing combined with sintering [14,15]. However, the structuration achieved by those methods is generally determined by the specificities inherent to the processing method and cannot be easily controlled and tuned by the engineer. Recently, magnetically assisted slip-casting (MASC) under high strength magnetic fields (10-15 T) has been adapted to ultra-low magnetic fields (below 50 mT) to create alumina ceramics with local and programmable grain orientations [16]. This orientation is creating a texture thanks to the well-defined crystallography of the basal plane of its anisotropic grains [16]. The ceramic specimens produced exhibited controllable and unusual textures, exemplified by layered and periodic orientations of anisotropic grains at defined angles, according to a pre-defined step variation of the applied magnetic field. Consolidation of the particles assembly after MASC to yield a ceramic was achieved by conducting the magnetic orientation of anisotropic magnetized microplatelets in a matrix of nanoparticles. During sintering, templated thermal grain growth (TGG) occurred, ultimately leading to the epitaxial growth of the microplatelets into larger grains, while retaining their orientation [17]. However, the final grain sizes achieved were large, with an average of 10 µm in diameter and 2 µm in thickness [16]. Since the mechanical and functional properties of ceramics depend on the size of their grains [18], there is a need to explore fabrication strategies where the texture, or grain orientation, can be controlled and tuned along with the grain size.

Grain dimensions in ceramics bodies are largely determined by the sintering process. In conventional sintering, temperature-driven diffusion leads to grain growth [19]. Lowering the sintering temperature and the sintering time as well as electrically-driven sintering processes such as spark plasma sintering [20], microwave sintering [21] or two step sintering [22], can yield smaller grain sizes. However, those methods act on the grain growth by slowing the grain boundary diffusion and are therefore not compatible with the TGG process. Indeed, during TGG, the abundant matrix nanoparticles are expected to merge with the microplatelets by diffusion-driven epitaxial growth [23]. In TGG, an appropriate method to modify the grain growth is to use ions in doping amounts. Ions such as $Si^{4+}$, $Mg^{2+}$, $Ca^{2+}$, act as flux and promote the grain growth along certain planes of the microplatelets. However, the addition of these dopants leads to large micrometric grains [12,17,24,25].

Submicrometric ceramics have been successfully realized by employing nanopowders as starting materials, while carefully controlling the sintering process. For example, nanocrystalline $MgAl_2O_4$ spinels were obtained starting with grains below 50 nm *via* an unusual negative Hall-Petch effect [26]. Submicrometric alumina could be prepared starting with nanospheres [27] or using hot isostatic pressing on nanopowders [28]. The materials resulting from these strategies exhibit enhanced but isotropic properties due to uniform and isotropic grain shapes. To combine controlled grain dimensions with anisotropy, the MASC-TGG process could use anisotropic microplatelets of various dimensions as starting powders.

In this study, we explore how MASC-TGG of alumina can be adapted to anisotropic microplatelets of various dimensions to yield textured ceramics with tunable grain size and orientation. Four types of microplatelets with differing lengths and aspect ratios were selected and magnetically functionalized. First, the magnetic response and the alignment characteristics of each microplatelet type are characterized in dilute conditions. Then, based on this knowledge, magnetically assisted slip casting is applied to slurries of microplatelets and nanoparticles to create anisotropic assemblies of particles in predefined orientations. Those assemblies are then consolidated into ceramics *via* TGG. The samples produced are characterized in terms of grain orientation, referred to as the texture, grain misorientation, size and overall porosity. Finally, comparing nanometric and micrometric mechanical properties, we discuss the potential benefits of building such textured materials for different applications, in particular structural particulate filters, bioceramics and ceramic preforms for reinforced composites.

## 2| EXPERIMENTAL PROCEDURE
### 2.1 | Magnetization of alumina platelets

Alumina microplatelets ($Al_2O_3$ pl, Kinsei Matec Co., Japan) of various dimensions (**Table 1**) were magnetized following an established procedure [29]. Typically, the microplatelets were suspended in a large volume of deionized water. After addition of superparamagnetic iron oxide particles (SPIONs, EMG-705, 10 nm diameter, Ferrotec) at a volume fraction with respect to alumina varying with the microplatelet type, the mixture was left stirring until all $Fe_3O_4$ nanoparticles adsorbed on the surface of the platelets. The functionalized powders were then filtered, rinsed with water and ethanol, and dried at 55°C under vacuum overnight. After functionalization, all powders have a brownish appearance, denoting successful magnetization. The concentration of SPIONs was calculated to have similar surface coverage of iron oxide of 21% [30].

**Table 1:** Dimensions of the alumina platelets: diameter ($L_i$), thickness ($l_i$) and aspect ratio $s_i$ and magnetic susceptibility $\chi$.

| Name | 02025 | 05070 | 07070 | 10030 |
| $L_i$ (μm) | 2.5 ± 0.5 | 4.05 ± 0.9 | 5.9 ± 1.4 | 6.39 ± 1.4 |
| $l_i$ (μm) | 0.25 ± 0.13 | 0.7 ± 0.2 | 0.7 ± 0.2 | 0.3 ± 0.26 |
| $s_i$ | 10.2 ± 1.4 | 5.8 ± 1 | 8.5 ± 1.4 | 21.3 ± 3.4 |
| $\chi_i$ | 7.28 | 16.96 | 23.28 | 16.76 |

## 2.2 | Magnetic response in dilute suspensions

*Suspension preparation:* Dried magnetized platelets were dispersed at 0.025 vol% into aqueous solutions of polyvinylpyrrolidone (PVP, 360'000 MW, Sigma-Aldrich, China) at 2, 5 and 10 wt%. The suspensions were mechanically mixed until homogeneous.

*Viscosity of the PVP solutions:* The shear profiles of the PVP solutions were recorded using a rheometer (Bohlin Gemini II, Germany) with a cone-plate set-up. The viscosities used for the calculations were estimated using the values at 0.01 s$^{-1}$ (see Supplementary Information (SI) **Figure S1** for the rheological curves).

*Magnetic set-up:* To visualize the platelet's response, a drop of suspension was deposited onto a glass slide on an inverted optical microscope (Ephiphot, Nikon) and images were recorded with the camera (Lumenra Infinity 1) and the software Infinity Analyze. The magnetic orientation set-up was positioned 3 cm above the drop and comprised a Neodymium magnet attached to a motor (940D271, RS components, Singapore) powered with 15 V (power supply Maisheng MS-3010D) (see **Figure S2** in SI for the set-up).

*Characterization of the alignment:* Image J (NIH, USA) was used to characterize the recorded images. In each image, the pixel coverage of a total of 30 platelets minimum was measured. The orientation angle $\theta$ with respect to the vertical was calculated using:

$$\theta = \cos^{-1}\left(\frac{P\prime}{P}\right),$$

where *P'* is the pixel count of the area of the microplatelet after alignment, and *P* is the pixel count of the same platelet before alignment [31]. Angle distributions combining results from all the images were obtained using Matlab and the misalignment was determined using the full width at half maximum (FWHM) of those distributions.

The orientation factor *f* was calculated using Herman's formula [32]:

$$f = \frac{3\overline{\cos^2 \theta} - 1}{2},$$

with $\overline{\cos^2 \theta}$ the average over all the angles recorded in an image.

The time of alignment was measured on recorded videos as the time at which all platelets aligned in the same direction upon application of the rotating magnetic field.

## 2.3 | Magnetically assisted slip casting

*Slurry preparation:* Functionalized alumina microplatelets were dispersed with alumina nanoparticles of ~120 nm diameter ($Al_2O_3$ np, Aerodisp W440, Evonik, Germany) at a microplatelets- nanoparticles ratio of 1:2. Efficient dispersion was ensured using acidic pH ~3-5 and mechanical mixing.

*Magnetically assisted slip casting:* The slurries were casted into molds made of 1 cm plastic tubes attached to a flat porous gypsum substrate (Ceramix, Germany), prepared according to the manufacturer's recommendations. The molds were placed on a magnetic alignment set-up for either horizontal or vertical orientation. The horizontal alignment set-up consisted of a sample attached on a standing-up rotating motor (RS component) placed at ~ 2 cm from the center of a strong Neodymium magnet. The magnetic field at the place of the sample was higher than 0.5 mT and the rotating speed of the motor ~1 Hz (see **Figure S3A** in SI for the set-up). The vertical alignment set-up consisted of a Neodymium magnet attached onto a motor in the horizontal position, whereas the samples were immobile. Similarly, the magnetic field at the area of the sample was higher than 0.5 mT and the rotating speed of the motor ~1 Hz (see **Figure S3B** in SI for the set-up). The magnetic field was maintained during the entire duration of the slip-casting and the removal of the water from the suspension through the pores of the mold.

*Green body characterization:* After the casting, the samples were dried at 30°C overnight (Binder VD 53, Fischer Scientific Pte Ltd, Singapore) before unmolding. Microscopic characterization of the green body was done by fracturing the pieces, mounting them onto a metallic stub and coating them with ~10 nm of platinum. Images were taken with a Scanning Electron Microscope (SEM, JEOL 76000F, Japan).

## 2.4 | Templated grain growth

*Sintering:* The green bodies were sintered in air for 2 h at 1600 °C (Nabertherm, Switzerland). The heating rate was set at 2.5 °C/min with an intermediate plateau of 5 h at 500 °C.

*Characterization of the sintered ceramics:* Cross-sections of the sintered ceramics were polished with SiC paper, coated with gold and observed by SEM (JSM-7600F, JEOL, Japan). The densities were determined using Archimedes' principle after overnight impregnation in ethanol, using the following formula:

$$\rho = \rho_e \times \frac{W_A}{W_{IA}-W_{IW}},$$

with $\rho$ the density of the ceramic, $\rho_e$ the density of ethanol, $W_A$ the dried weight, $W_{IA}$ the weight of the impregnated sample measured in air and $W_{IW}$ the weight of the impregnated measured in ethanol. The closed porosity $CP$ was calculated using:

$$CP = \frac{\rho_T - \rho}{\rho_T} \times 100 - \frac{W_{IA} - W_A}{W_{IA} - W_{IW}} \times 100,$$

with $\rho_T$ the theoretical density of alumina.

The anisotropic shrinkage was measured on horizontally and vertically aligned disc samples by measuring the height and diameter before and after sintering. The measurements were averaged over 3 to 5 samples.

The grain sizes after sintering were measured on the SEM images using the software Image J (NIH, USA) and averaged over 20 to 50 grains per image.

### 2.5 | Mechanical properties

*Sample preparation:* Sintered specimens were hot mounted in resin, mirror polished (AP-D and OPS colloidal solutions, Struers, Denmark), and ultrasonicated to remove debris.

*Nanoscale properties:* Twenty indents spaced apart with 50 µm were done on each sample using a Berkovitch tip mounted on a nanoindenter (G200, KLA Tencor, US). The tests were run under load-controlled conditions, with a loading rate of 1 mN/s, a maximum load of 200 mN and a thermal drift of 0.1 nm/s. The peak hold time was set to 10 s and the Poisson's ratio assumed to be 0.22. The data were analyzed using Oliver-Pharr. The reliability was checked using Weibull analysis (m>3).

*Microscale properties:* Ten indents were made at a loading force of 500 g and dwell time of 10 s using a Vicker's indenter (Future-Tech microhardness tester FM-300E). The indented areas were observed under optical microscopy. Image J was used to measure the indented area and calculate the hardness.

*Post-mortem fractography:* Indented and fractured samples were observed under SEM after 10 nm platinum coating.

### 3| RESULTS AND DISCUSSION

### 3.1 | Magnetic orientation as a function of microplatelet's dimensions

Using ultra low magnetic fields to orient anisotropic particles requires magnetically responsive particles. It is known that the magnetic response, in particular the ultra-high magnetic response (UHMR), is not only influenced by the field applied, but also by the dimensions of the particles to orient [29]. Indeed, when the dimensions of those particles are too little, the thermal energy predominates over the magnetic energy, hindering the alignment. Similarly, when the dimensions are too large, the gravitational forces dominate and tend to align the particles horizontally. To use MASC for magnetically responsive particles of various dimensions, we first need to characterize their magnetic response (**Figure 1**).

The microparticles considered are alumina microplatelets with diameters varying between ~2 to 10 µm (**Table 1**). Those microplatelets can be produced in bulk scale through molten salt synthesis or hydrothermal methods [33] and are available commercially. This makes them interesting for bulk and large-scale applications. Alumina microplatelets can be functionalized by physically adsorbing charged superparamagnetic iron oxide nanoparticles (SPIONs) on their surface [30,34] (**Figure 1A**). Orientation and biaxial alignment of the obtained magnetized microplatelets is achieved by applying a rotating ultra-low magnetic field $\vec{H}_0$ spinning at a frequency $\omega$ [34]. Indeed, under magnetic fields rotating above a critical frequency $\omega_c$, magnetically-responsive anisotropic particles are found to transition from a rotational motion synchronized with the field to a phase-ejected mode where they align in the plane of rotation of the field [34]. This phase-ejection and consequent biaxial alignment is due to the energetically unfavorable viscous drag generated during the rotational motion. In the phase-ejection mode, this viscous drag is suppressed hence is more energetically favorable [34].

Suspended in a low-viscosity background fluid (~8 mPa.s) at a dilute concentration of ~0.02 vol%, largely inferior to the concentration of self-assembly into nematic order (Onsager concentration $\nu_{I-N}$ in **Table 2,** calculation detailed in SI), the microplatelets used in this study exhibited the expected behavior (**Figure 1B**). Indeed, in absence of magnetic field, the largest platelets were found to sediment and adopt a horizontal alignment whereas the smaller particles remained suspended. Some orientation in absence of magnetic field could be observed with the smaller platelets with a high orientation factor *f* instead of the expected value of zero (**Figure 1C,** empty circles). This can be presumably due to the flow induced during the deposition of the drop. After application of a ultra-low magnetic field of ~0.45 mT rotating at a frequency of ~8 Hz, pronounced vertical biaxial alignment was recorded, with an orientation factor *f* close to 1 (**Figure 1B,C,** full circles). The rotation frequency was chosen so as to be well above the theoretical critical frequency required for the alignment (**Table 2**, calculation details in SI).

The magnetic response of the platelets displayed similar behavior in background liquids of increasing viscosities (**Figure 1D-F**). A higher magnetic orientation factor *f* and smaller misalignment were observed for platelets with larger diameters and in solutions of higher viscosity. Indeed, for alumina, the critical size at which ultra-high magnetic response occurs is close to 10 µm [29]. At this diameter, the magnetic energy largely overcomes the thermal agitation and gravity. Also, a larger background viscosity induces a larger viscous force, leading to preferential alignment in the plane of rotation of the magnet. However, direct comparison of the platelets response to the magnetic field should take into account both the diameter and the aspect ratio. Although the platelet's length had the most incidence on the alignment quality, the time to align generally increased with the aspect ratio, as well as with the background viscosity (**Figure 1F**). Indeed, a large aspect ratio, hence anisotropy, generates a large viscous drag so that it becomes more favorable for the platelet to align [35].

We have thus verified that the magnetically functionalized microplatelets used in this study could be directed vertically using an ultra-low external rotating magnetic field. Since vertical alignment is the least favorable orientation, it can be anticipated that any other orientation could be achieved as well [36]. In the following, we use those functionalized microplatelets in combination to MASC and TGG to fabricate textured ceramics.

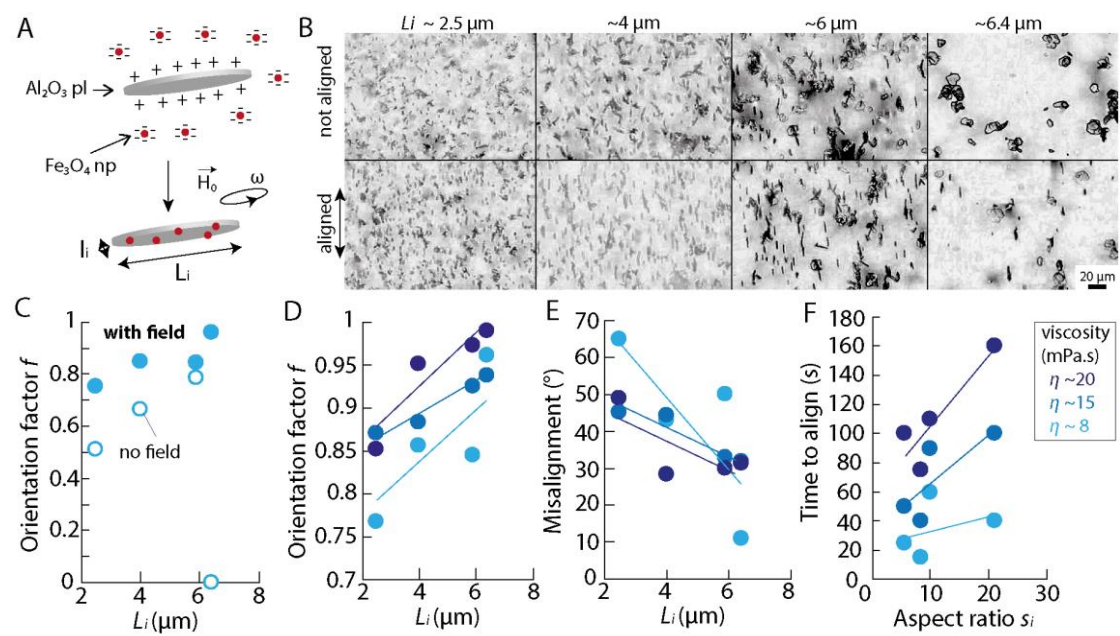

**FIGURE 1.** **(A)** Schematics of the magnetization of $Al_2O_3$ platelets (pl) of diameter $L_i$ and thickness $l_i$ using $Fe_3O_4$ nanoparticles (np) of 10 nm diameter. A magnetic field $H_0$ rotating at a frequency $\omega$ aligns the platelets. **(B)** Optical images of the $Al_2O_3$ platelets suspended in 2 wt% PVP before and after vertical biaxial magnetic alignment. **(C)** Herman's orientation factor $f$ as a function of $L_i$ before and after alignment in background liquid of viscosity $\eta \sim 8\ mPa.s$, using $H_0 = 0.45\ mT$ and $\omega = 8\ Hz$. **(D,E)** Orientation factor $f$ and misalignment angles as functions of $L_i$ after alignment for increasing background viscosity. The lines are guides to the eyes. **(F)** Time to align as a function of the aspect ratio $s_i$. The lines are guides to the eyes.

**Table 2:** Theoretical calculations of the concentration at which spontaneous alignment is expected to occur ($\nu_{I-N}$) and critical alignment frequencies $\omega_c$ at different background liquid viscosities for the alumina platelets of diameter ($L_i$) and aspect ratio $s_i$ and under a magnetic field of 0.5 mT.

| $L_i$ (μm) | 2.5 | 4.05 | 5.9 | 6.39 |
|---|---|---|---|---|
| $s_i$ | 10.2 | 5.8 | 8.5 | 21.3 |
| $\nu_{I-N}$ (vol%) | 52.3 | 92 | 22.9 | 63 |
| $\omega_c$ (Hz) at $\eta = 8\ mPa.s$ | 0.023 | 0.037 | 0.034 | 0.020 |
| $\omega_c$ (Hz) at $\eta = 15\ mPa.s$ | 0.012 | 0.020 | 0.019 | 0.011 |
| $\omega_c$ (Hz) at $\eta = 20\ mPa.s$ | 0.009 | 0.015 | 0.014 | 0.008 |

### 3.2 | Magnetic alignment after slip casting of the bimodal slurries

To verify the suitability of using microplatelets of various dimensions for MASC-TGG, the magnetized alumina platelets prepared as described were dispersed in a suspension of alumina nanoparticles to form a bimodal slurry. Since the viscosity of suspensions of particles increases as the particle size decreases and as their anisotropy increases [37], the ratio nanoparticles: templates used was of 2:1 to maintain fluidity [30]. At these concentrations, all magnetic platelets could be efficiently dispersed and did not exhibit sedimentation in 2 to 3 mm-thick samples. The magnetically driven orientation of the alumina platelets was maintained thanks to the low viscosity of the well-dispersed slurry of nanoparticles, below 100 mPa.s. According to the experiments with the PVP solutions (**Figure 1D-F**), we expect that an increase in viscosity will improve the alignment. However, the alignment time is also expected to increase, proportional to the viscosity [35] (**Figure 2A,** see SI for details on the fits). We estimate that the time required to align the platelets under a field of 0.45 mT and in a fluid of 100 mPa.s is of 15 minutes maximum for the largest set of particles. This alignment time is inferior to the time the magnetic field was applied in the experimental set-up, of 3 hour minimum. Indeed, we conducted the magnetic orientation during the entire time of the slip casting. Besides, a much shorter time has been reported earlier, suggesting that the fluid directly surrounding the microplatelets, *i.e.* water, is the fluid to consider [16]. This implies that magnetic orientation would occur without hindrance in the bimodal microplatelets-nanoparticles system.

During slip-casting, the fluid from the suspension is slowly removed through the porous mold by capillary forces. Those capillary forces, and the resulting hydrodynamic flow, have been reported to disturb the alignment of particles [36]. In homogeneous and well-dispersed suspensions of microplatelets and nanoparticles, it is expected that the nanoparticles get trapped and concentrated in between aligned platelets, maintaining their orientation after drying (**Figure 2B**). Inspecting the assembly after MASC, alumina nanoparticles were indeed highly packed between aligned template platelets (**Figure 2C**). The overall magnetic alignment within the green body could be visualized by the observation of fractured cross-sections. During fracture, some platelets get pulled-out, leaving black lines at their original position (**Figure 2D**).

The alignment of magnetized platelets with different dimensions occurs during MASC. The grain growth by TGG and the texture in the sintered ceramics are characterised in the next paragraph.

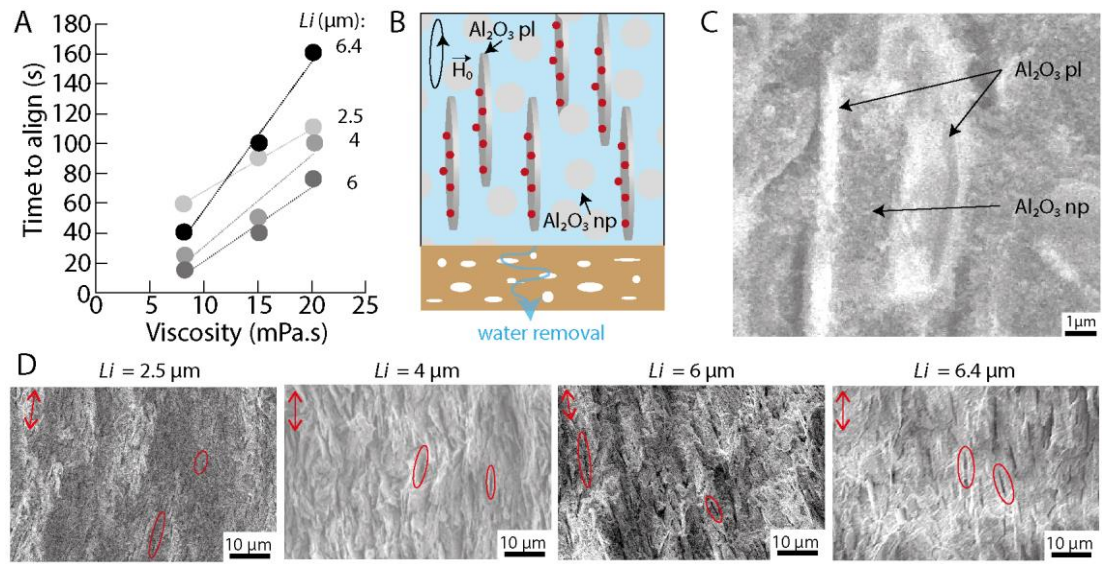

**FIGURE 2. (A)** Time to align as a function of the background viscosity and platelet diameter $L_i$. The lines are linear fits. **(B)** Schematics of the magnetically assisted slip-casting process showing $Al_2O_3$ np "trapped" between vertically oriented $Al_2O_3$ pl. **(C)** Electron micrograph of a dried green body made with platelets of diameter $L_i$ = 6.4 and vertical orientation showing the local assembly of particles. **(D)** Micrographs of fractured cross-sections of dried green bodies. Position and orientations of some microplatelets are circled in red.

**3.3 | Templated grain growth and texturation**

After magnetic slip-casting and drying, the green bodies were sintered in air at 1600°C for 2 hours. Due to the magnetic alignment, the shrinkage was anisotropic with a higher shrinkage perpendicular to the microplatelet's long axis (**Figure 3A**). Indeed, in this perpendicular direction, the shrinkage was found to be independent to the initial aspect ratio of the microplatelets. This can be explained by their similar thickness, between 0.25 and 0.7 µm, but also by the high concentration and packing of nanoparticles along this direction (**Figure 2C**). However, the shrinkage parallel to the long axis of the platelets is correlated to the initial aspect ratio. Although it could have been anticipated that larger aspect ratio would hinder the shrinkage the most due to higher modulus of the platelets in this direction and the difficulty of rearrangement, the opposite trend was observed. It can be argued that larger aspect ratio actually enabled more contact between the template microplatelets and the nanoparticles, creating more necking and sintering opportunities.

After sintering, the final diameter $L_f$ of the grains varied linearly with the initial diameter $L_i$. This followed the theoretical expectations from the theory of TGG [23,30] (**Figure 3B,** see details of the calculation in SI). The final grains varied from ~4 to 7 µm in diameter. However, the grain thicknesses remained the same for all initial platelet diameters, of ~0.5 µm. The sintering process also maintained the alignment of the templates during TGG so that the final ceramics exhibited little misalignment between only 10 and 20° (**Figure 3C,D**). This misalignment is lower than the one measured in the PVP solution where it varied globally between 30 and 60°. This follows the expected trend that a higher background viscosity improves the alignment. The grains obtained after sintering had an aspect ratios $s_f$ varying between 5 and 15 (**Figure 3E**). This is a larger range of aspect ratios than those obtained in other works, reported between 8 and 10 or around 5, and for smaller grain sizes as compared to the 10 to 30 µm diameters in the literature [12,16]. This makes the ceramic described herein promising for enhanced mechanical and functional properties.

Finally, the density of the MASC-TGG ceramics was also dependent on the grain size. The larger the final grain size, the lower the density (**Figure 3E**). It can be hypothesized that anisotropic TGG hindered the densification. Nevertheless, the close porosity was found close to zero, which is a key feature for filtering applications or to use the ceramics as preforms for highly reinforced anisotropic composites.

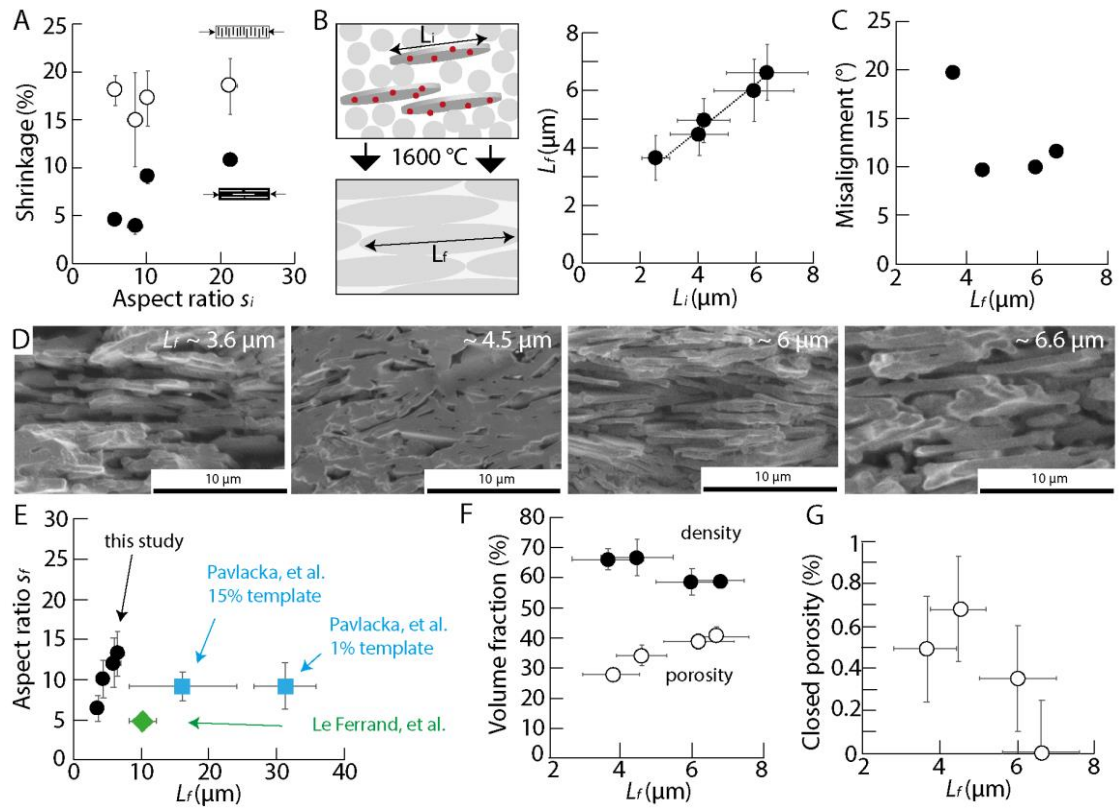

**FIGURE 3. (A)** Sintering shrinkage as a function of the aspect ratio $s_i$ and microplatelet's orientation: perpendicular to the alignment, white; parallel, black. **(B)** Schematics of the TGG process and final grain diameter $L_f$ as a function of the initial microplatelet diameter $L_i$. The dotted line represents calculated predictions. **(C)** Misalignment angles in the sintered ceramics as a function of $L_f$. **(D)** Electron micrographs of the cross-sections of aligned ceramics. **(E)** Aspect ratio $s_i$ of the sintered pieces as a function of their grain diameter $L_f$ and comparison with the literature [12,16]. Density, open **(F)** and closed **(G)** porosity as a function of $L_f$.

### 3.4 | Properties and potential applications

The mechanical properties of the porous textured ceramics achieved were characterised at the nanometric and micrometric scales for vertical and horizontal orientations of their grains (**Figure 4**).

First, nanoindentation provided information regarding the hardness and the elastic modulus at the nanoscale (**Figure 4A-C**). The mechanical properties were found to drastically decrease with the grain diameters, as expected from Griffith. This effect is likely to be amplified by the increase in porosity with $L_f$. Mild anisotropy in hardness could be recorded for the ceramic with $L_f$ < 6 µm, whereas the elastic moduli did not vary with the orientation. However, it is the horizontally aligned samples that exhibited the higher hardness, which is counter intuitive but could be explained by the fact that at the nanometric scale, there are less interfaces encountered in the horizontal configuration. However, as expected, the trend is opposite for microindentation using a Vickers indenter, where vertically aligned ceramics displayed a higher hardness, ca. 15 GPa, as compared to the horizontal orientation, ca. 10 GPa (**Figure 4D**). Again, the hardness values were highly correlated with the density of the ceramics and thus decreased for grains of larger anisotropy and diameter. In the denser specimens, the difference between vertical and horizontal orientations on the mechanics was clearly visible in fractography (**Figure 4E**). Indentation on the horizontal surface led to low hardness through removal of the anisotropic grains, leaving a large imprint, whereas this was not observed on the vertical samples, where only a few platelets were sheared away. Similar mechanisms have been observed in another study [16].

Finally, there is also interest in fabricating textured ceramics to implement extrinsic toughening mechanisms in ceramics and their composites [38]. Through the control of the orientation of grains, cracks can be directed. Hence, it is possible to generate extensive crack twisting and deflection, as cracks travel along the weaker grain interfaces. As a qualitative example of such extensive crack deflection and delamination, **Figure 4F** depicts naturally-occurring cracks after compressing a horizontally aligned sample with grain diameter $L_f$ ~6 µm. To promote crack deflection and toughening in ceramics, a high aspect ratio, high density, and large interfacial strength are required [39]. Thanks to the low closed porosity of the fabricated textured ceramics, the materials developed in this work offer promising opportunities to be used as preforms for reinforced composite materials. In addition, the possibility to tune and control independently both grain orientation and size could also be exploited for diverse applications in porous ceramics, for filtering, catalysis, or tissue engineering.

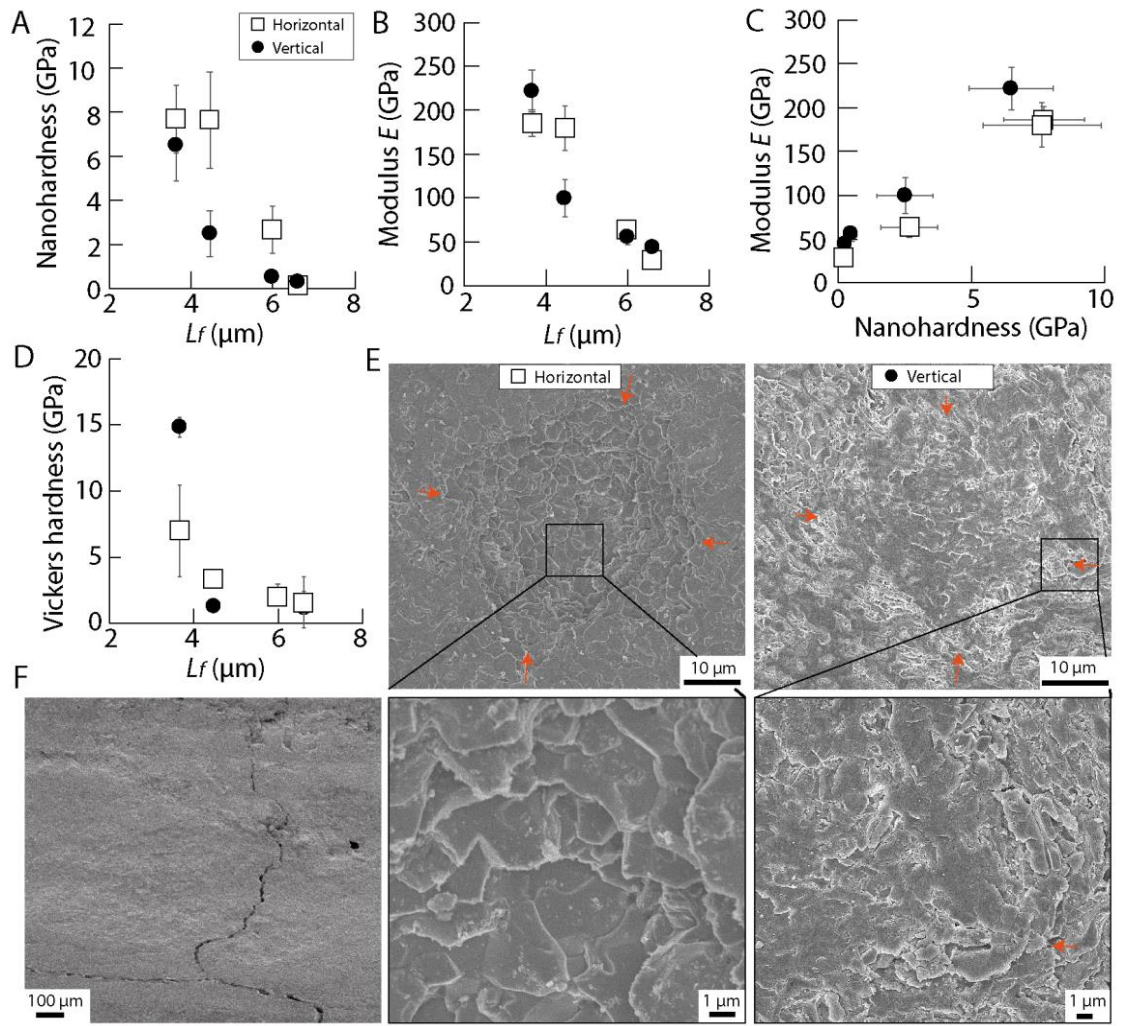

**FIGURE 4.** Nanohardness **(A)** and modulus **(B)** measured by nanoindentation as functions of the grain diameter $L_f$ and orientation: vertical with respect to the indenting direction, black; horizontal, white. **(C)** Modulus as a function of the nanohardness. **(D)** Microscopic Vickers hardness as a function of $L_f$ and grain orientation and **(E)** *post-mortem* micrographs of the indents on oriented ceramics with $L_f \sim 4~\mu m$. **(F)** Tortuous crack path in a ceramic with $L_f \sim 6.6~\mu m$.

## 4 | CONCLUSIONS

In this study, we have applied magnetically assisted slip casting and templated grain growth to alumina microplatelets of various dimensions to prepare ceramics with independently controlled grain orientations and grain sizes. It was found that for the microplatelets chosen, magnetic orientation could be achieved and that the grain sizes in the sintered bodies were lower than those achieved in other work, and with higher aspect ratios. The mechanical properties exhibited interesting crack deflection and anisotropy but were predominantly controlled by the open porosity that remained in the textured ceramics. To circumvent the porosity problem that is difficult to decouple from the microplatelets' dimensions, the ceramics could be infiltrated by a secondary phase, thanks to the minimal closed porosity. Also, the ratio of microplatelets to nanoparticles could be optimized for each microplatelet type. Although the final grain sizes are still in the micrometric range, the thickness remained submicrometric, which is a key feature for enhanced mechanical properties when a load is applied along the thickness. Finally, although a limitation in the MASC-TGG process is that nanopowders would require ultra-high magnetic field strengths to overcome thermal agitation, the microstructures presented here are unique, scalable, and largely tunable for the processing of composites and ceramics.

## ACKNOWLEDGMENTS

The authors are thankful to Kinsei, Japan, for providing the alumina microplatelets. The authors acknowledge financial support from Nanyang Technological University, Singapore (Start-Up grant).

## SUPPORTING INFORMATION

Supporting information is available at the attached PDF.

## DECLARATION OF COMPETING INTEREST

The authors declare that they have no known competing financial interests or personal relationships that could have appeared to influence the work reported in this paper.

## ORCID

Hortense Le Ferrand. Orcid.org/0000-0003-3017-9403

## REFERENCES

[1]  C. Chen, B.R. Müller, C. Prinz, J. Stroh, I. Feldmann, G. Bruno, The correlation between porosity characteristics and the crystallographic texture in extruded stabilized aluminium titanate for diesel particulate filter applications, J. Eur. Ceram. Soc. 40 (2020) 1592–1601. doi:10.1016/j.jeurceramsoc.2019.11.076.

[2]  J. Billaud, F. Bouville, T. Magrini, C. Villevieille, A.R. Studart, Magnetically aligned graphite electrodes for high-rate performance Li-ion batteries, Nat. Energy. 1 (2016) 1–7. doi:10.1038/nenergy.2016.97.

[3]  M.R. Sommer, R.M. Erb, A.R. Studart, Injectable materials with magnetically controlled anisotropic porosity, ACS Appl. Mater. Interfaces. 4 (2012) 5086–5091. doi:10.1021/am301500z.

[4]     T.P. Niebel, F. Bouville, D. Kokkinis, A.R. Studart, Role of the polymer phase in the mechanics of nacre-like composites, J. Mech. Phys. Solids. 96 (2016) 133–146. doi:10.1016/j.jmps.2016.06.011.

[5]     T. Magrini, F. Bouville, A. Lauria, H. Le Ferrand, T.P. Niebel, A.R. Studart, Transparent and tough bulk composites inspired by nacre, Nat. Commun. 10 (2019) 1–10. doi:10.1038/s41467-019-10829-2.

[6]     A. Bale, R. Rouffaud, A.C. Hladky-Hennion, P. Marchet, F. Levassort, Modeling the electroelastic moduli of porous textured piezoceramics, IEEE Trans. Ultrason. Ferroelectr. Freq. Control. 66 (2019) 949–957. doi:10.1109/TUFFC.2019.2898519.

[7]     O. Rozenbaum, D. De Sousa Meneses, P. Echegut, Texture and porosity effects on the thermal radiative behavior of alumina ceramics, Int. J. Thermophys. 30 (2009) 580–590. doi:10.1007/s10765-008-0510-1.

[8]     R. Shimonishi, M. Hagiwara, S. Fujihara, Fabrication of highly textured $Ca_3Co_4O_9$ ceramics with controlled density and high thermoelectric power factors, J. Eur. Ceram. Soc. 40 (2020) 1338–1343. doi:10.1016/j.jeurceramsoc.2019.11.077.

[9]     P.I.B.G.B. Pelissari, V.C. Pandolfelli, D. Carnelli, F. Bouville, Refractory interphase and its role on the mechanical properties of boron containing nacre-like ceramic, J. Eur. Ceram. Soc. 40 (2020) 165–172. doi:10.1016/j.jeurceramsoc.2019.08.034.

[10]    H. Le Ferrand, F. Bouville, T.P. Niebel, A.R. Studart, Magnetically assisted slip casting of bioinspired heterogeneous composites, Nat. Mater. 14 (2015) 1172–1179. doi:10.1038/nmat4419.

[11]    I. Nelson, S.E. Naleway, Intrinsic and extrinsic control of freeze casting, Integr. Med. Res. 8 (2019) 2372–2385. doi:10.1016/j.jmrt.2018.11.011.

[12]    R.J. Pavlacka, G.L. Messing, Processing and mechanical response of highly textured Al2O3, J. Eur. Ceram. Soc. 30 (2010) 2917–2925. doi:10.1016/j.jeurceramsoc.2010.02.009.

[13]    H. Le Ferrand, Manuscript_MAScreview_JEurCeramSoc, Submitted. (n.d.).

[14]    Z. Yang, J. Yu, C. Li, Y. Zhong, K. Deng, Z. Ren, Q. Wang, Y. Dai, H. Wang, Effect of β-$Si_3N_4$ Initial Powder Size on Texture Development of Porous $Si_3N_4$ Ceramics Prepared by Gel-Casting in a Magnetic Field, Trans. Indian Ceram. Soc. 75 (2016) 256–262. doi:10.1080/0371750X.2016.1202143.

[15]    S. Honda, S. Hashimoto, S. Iwata, Y. Iwamoto, Anisotropic properties of highly textured porous alumina formed from platelets, Ceram. Int. 42 (2016) 1453–1458. doi:10.1016/j.ceramint.2015.09.090.

[16] H. Le Ferrand, F. Bouville, Processing of dense bioinspired ceramics with deliberate microstructure, J. Am. Ceram. Soc. (2019) 1–11. doi:10.1111/jace.16656.

[17] M.M. Seabaugh, I.H. Kerscht, G.L. Messing, Texture Development by Templated Grain Growth in Liquid-Phase-Sintered α-Alumina, J. Am. Ceram. Soc. 80 (2005) 1181–1188. doi:10.1111/j.1151-2916.1997.tb02961.x.

[18] D.P.H. Hasselman, Griffith Flaws and the Effect of Porosity on Tensile Strength of Brittle Ceramics, J. Am. Ceram. Soc. 56 (1969) 457. doi:10.1111/j.1151-2916.1973.tb12533.x.

[19] C. Manière, S. Chan, G. Lee, J. McKittrick, E.A. Olevsky, Sintering dilatometry based grain growth assessment, Results Phys. 10 (2018) 91–93. doi:10.1016/j.rinp.2018.05.014.

[20] M.S. Boldin, A.A. Popov, E.A. Lantsev, A. V. Nokhrin, V.N. Chuvil'Deev, Investigation of the kinetics of spark plasma sintering of alumina. Part 2. Intermediate and final stages of sintering, IOP Conf. Ser. Mater. Sci. Eng. 558 (2019) 012006. doi:10.1088/1757-899X/558/1/012006.

[21] O. Zgalat-Lozynskyy, A. Ragulya, Densification Kinetics and Structural Evolution During Microwave and Pressureless Sintering of 15 nm Titanium Nitride Powder, Nanoscale Res. Lett. 11 (2016) 1–9. doi:10.1186/s11671-016-1316-x.

[22] Z. Razavi Hesabi, M. Haghighatzadeh, M. Mazaheri, D. Galusek, S.K. Sadrnezhaad, Suppression of grain growth in sub-micrometer alumina via two-step sintering method, J. Eur. Ceram. Soc. 29 (2009) 1371–1377. doi:10.1016/j.jeurceramsoc.2008.08.027.

[23] E. Suvaci, K.S. Oh, G.L. Messing, Kinetics of template growth in alumina during the process of templated grain growth (TGG), Acta Mater. 49 (2001) 2075–2081. doi:10.1016/S1359-6454(01)00105-7.

[24] S.H. Hong, G.L. Messing, Development of textured mullite by templated grain growth, J. Am. Ceram. Soc. 82 (1999) 867–872. doi:10.1111/j.1151-2916.1999.tb01847.x.

[25] Y. Chang, S. Poterala, D. Yener, G.L. Messing, Fabrication of highly textured fine-grained α-alumina by templated grain growth of nanoscale precursors, J. Am. Ceram. Soc. 96 (2013) 1390–1397. doi:10.1111/jace.12286.

[26] H. Ryou, J.W. Drazin, K.J. Wahl, S.B. Qadri, E.P. Gorzkowski, B.N. Feigelson, J.A. Wollmershauser, Below the Hall-Petch Limit in Nanocrystalline Ceramics, ACS Nano. 12 (2018) 3083–3094. doi:10.1021/acsnano.7b07380.

[27] Z. Sun, B. Li, P. Hu, F. Ding, F. Yuan, Alumina ceramics with uniform grains

prepared from Al2O3 nanospheres, J. Alloys Compd. 688 (2016) 933–938. doi:10.1016/j.jallcom.2016.07.122.

[28] P. Liu, H. Yi, G. Zhou, J. Zhang, S. Wang, HIP and pressureless sintering of transparent alumina shaped by magnetic field assisted slip casting, Opt. Mater. Express. 5 (2015) 441. doi:10.1364/ome.5.000441.

[29] R.M. Erb, R. Libanori, N. Rothfuchs, A.R. Studart, Composites reinforced in three dimensions by using low magnetic fields, Science (80-. ). 335 (2012) 199–204. doi:10.1126/science.1210822.

[30] H. Le Ferrand, Pressure-less processing of ceramics with deliberate elongated grain orientation and size, in: B. Li, S. Baker, H. Zhai, R. Soman, J. Li, S. Monteiro, F. Dong, R. Wang (Eds.), Adv. Powder Ceram. Mater., Springer Nature Switzerland, 2020: pp. 45–56.

[31] C. Liu, J. Cai, D. Gong, W. Zhang, D. Zhang, Fabrication and uniform alignment of nickel-coated diatomite flakes by rotating magnetic fields, J. Magn. Magn. Mater. 484 (2019) 472–477. doi:10.1016/j.jmmm.2019.04.011.

[32] J.L. White, J.E. Spruiell, The specification of orientation and its development in polymer processing, Polym. Eng. Sci. 23 (1983) 247–256. doi:10.1002/pen.760230503.

[33] Y. Chang, J. Wu, M. Zhang, E. Kupp, G.L. Messing, Molten salt synthesis of morphology controlled α-alumina platelets, Ceram. Int. 43 (2017) 12684–12688. doi:10.1016/j.ceramint.2017.06.150.

[34] R.M. Erb, J. Segmehl, M. Charilaou, J.F. Löffler, A.R. Studart, Non-linear alignment dynamics in suspensions of platelets under rotating magnetic fields, Soft Matter. 8 (2012) 7604–7609. doi:10.1039/c2sm25650a.

[35] D. Kokkinis, M. Schaffner, A.R. Studart, Multimaterial magnetically assisted 3D printing of composite materials, Nat. Commun. 45 (2015) 333–338. doi:10.1038/ncomms9643.

[36] H. Le Ferrand, F. Bouville, A.R. Studart, Design of textured multi-layered structures via magnetically assisted slip casting, Soft Matter. 15 (2019) 3886. doi:10.1039/c9sm00390h.

[37] S. Mueller, E.W. Llewellin, H.M. Mader, The rheology of suspensions of solid particles, Proc. R. Soc. A. 466 (2010) 1201–1228. doi:10.1007/BF01432034.

[38] W. Huang, D. Restrepo, J.Y. Jung, F.Y. Su, Z. Liu, R.O. Ritchie, J. McKittrick, P. Zavattieri, D. Kisailus, Multiscale Toughening Mechanisms in Biological Materials and Bioinspired Designs, Adv. Mater. 1901561 (2019) 1–37. doi:10.1002/adma.201901561.

[39] F. Barthelat, Designing nacre-like materials for simultaneous stiffness,

strength and toughness: Optimum materials, composition, microstructure and size, J. Mech. Phys. Solids. 73 (2014) 22–37. doi:10.1016/j.jmps.2014.08.008.